\documentclass[twocolumn,showpacs,superscriptaddress]{revtex4}
\usepackage{graphicx}
\usepackage{amsmath}

\begin{document}

\title{Gas-Liquid Coexistence in the Primitive Model for Water}
\author{Flavio Romano\footnote{Author to which correspondence should be 
addressed; electronic address {\tt flavio.romano@gmail.com}}}
\affiliation{Dipartimento di Fisica, Universit\`a di Roma {\em La Sapienza}, 
Piazzale A. Moro 2, 00185 Roma, Italy}
\author{Piero Tartaglia}
\affiliation{Dipartimento di Fisica and INFM-CNR-SMC, Universit\`a di Roma 
{\em La Sapienza}, Piazzale A. Moro 2, 00185 Roma, Italy} 
\author{Francesco Sciortino}
\affiliation{Dipartimento di Fisica and INFM-CNR-SOFT, Universit\`a di Roma 
{\em La Sapienza}, Piazzale A. Moro 2, 00185 Roma, Italy}

\begin{abstract}
We evaluate the location of the gas-liquid coexistence line and of the 
associated critical point for the primitive model for water (PMW), introduced 
by Kolafa and Nezbeda [J. Kolafa and I. Nezbeda, Mol. Phys. {\bf 61}, 161 
(1987)]. Besides being a simple model for a molecular network forming liquid, 
the PMW is representative of patchy proteins and novel colloidal particles 
interacting with localized directional short-range attractions. We show that 
the gas-liquid phase separation is metastable, i.e. it takes place in the 
region of the phase diagram where the crystal phase is thermodynamically 
favored, as in the case of particles interacting via short-range attractive 
spherical potentials. Differently from spherical potentials, we do not observe 
crystallization close to the critical point. The region of gas-liquid 
instability of this patchy model is significantly reduced as compared to 
equivalent models of spherically interacting particles, confirming the 
possibility of observing kinetic arrest in an homogeneous sample driven by 
bonding as opposed to packing.
\end{abstract}


\maketitle

\section{Introduction}
This article presents a detailed numerical study of the critical point and 
gas-liquid coexistence of a model introduced several years ago by Kolafa and 
Nezbeda~\cite{Kol87a} as a primitive model for water (PMW). The water molecule 
is described as a hard-sphere with four interaction sites, arranged on a 
tetrahedral geometry, which are meant to mimic the two hydrogen protons and 
the two oxygen lone-pairs of the water molecule. The PMW has been studied in 
detail in the past, since it is both a valid candidate for testing theories of 
bond association~\cite{Wer84a,Wer84b,Gho93a,Sea96a,Dud98a,Pee03a,Kal03a} as 
well as a model able to reproduce the thermodynamic anomalies of 
water~\cite{Kol87a,Nez89a,Nez90a,Veg98a,pwmnoi}. The Wertheim 
theory~\cite{Wer84a,Wer84b} has been carefully compared to numerical studies, 
suggesting a good agreement between theoretical predictions and numerical data 
in the temperature $T$ region where no significant ring formation is 
observed~\cite{Veg98a,pwmnoi}. More recently, the slow dynamics of this model 
has been studied using a newly developed code for event-driven dynamics of 
patchy particles~\cite{pwmnoi}. It has been shown that, at low density, there 
are indications of a gas-liquid phase separation. At intermediate density, the 
system can be cooled down to the smallest temperatures at which equilibration 
is feasible with present day computational facilities without any signature of 
phase separation. In this region, dynamics progressively slows down following 
an Arrhenius law, consistently with the expected behavior of strong 
network-forming liquids.

This primitive model, besides its interest as an elementary model for water, 
is also representative of a larger class of models which are nowadays receiving 
considerable interest, namely models of particles interacting with localized 
directional short-range attractions. This class of models includes, apart from 
network forming molecular systems, also 
proteins~\cite{Lom99a,Sea99c,Ker03a,doye} and newly designed colloidal 
particles~\cite{Man03a}. Recently, the phase diagram of patchy particles has 
been studied~\cite{bianchiprl} in the attempt to estimate the role of the
surface patchiness in dictating their dynamic and thermodynamic behavior. It 
has been suggested that the number of possible bonds $M$, as opposed to the 
fraction of surface with attractive interactions, is the key ingredient in 
determining the width of the unstable region of the phase 
diagram~\cite{Zac05a,Sci04bCPC,zaccalungo}. A study of the evolution of the 
critical point on decreasing $M$ shows a clear monotonic 
trend~\cite{bianchiprl} and, more importantly, in the direction of decreasing 
critical packing fraction. Such a trend is not observed in spherical 
potentials on decreasing the attraction range. Thus, in low-valence particle 
systems, a window of packing fraction $\phi$ values opens up in which it is 
possible to reach very low $T$ (and hence states with extremely long bond 
lifetimes) without encountering phase separation. This favors the 
establishment of a spanning network of long-living bonds, which in the 
colloidal community provides indication of gel formation but which, in the 
field of network forming liquids, would be rather classified as glass 
formation~\cite{silica}. 

As previously mentioned, we present here an accurate estimate of the location 
of the critical point for the PMW (for which $M=4$), based on finite-size 
scaling, and of the associated gas-liquid coexistence phase diagram. Comparing 
with the known fluid-crystal phase coexistence loci~\cite{Veg98a} we are able 
to show that the gas-liquid phase separation is metastable as compared to the 
crystal state. Despite its metastability, we have never observed 
crystallization, not even close to the critical point, an observation which 
could be of relevance in the case of the crystallization of patchy proteins or 
colloids. We also show that the number density $\rho$ of the liquid phase at 
coexistence is comparable to the crystal (diamond) density, much smaller than 
the characteristic liquid density observed in systems of particles interacting 
through spherically symmetric potentials. 

\begin{figure}[tb]
\includegraphics[width=8cm]{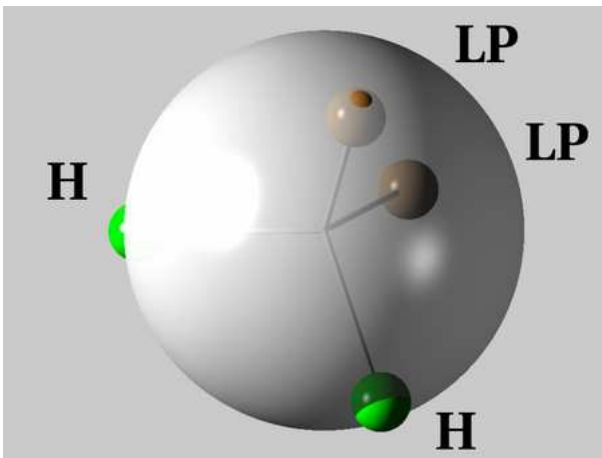}
\caption{Pictorial representation of the model studied. Each particle is 
modeled as an hard-core sphere (grey large sphere of diameter $\sigma$). 
The four interaction sites are located in a tetrahedral arrangement. Two of 
the sites (the H-sites, in green) are located on the surface whereas the 
remaining two sites (LP, in brown) are located inside the sphere, at distance 
$0.45 \sigma$ from the center. Only H-sites and L-sites on different particles 
interact with a square well interaction of depth $u_0$ and width 
$\delta=0.15\sigma$.}
\label{fig:m2}
\end{figure}

\section{The model}
In the PMW, each particle is composed by a hard sphere of diameter $\sigma$ 
and four additional sites arranged according to a tetrahedral geometry. 
Two of these sites, the proton sites H, are located on the surface of the hard 
sphere, i.e. at distance $0.5\sigma$ from the center of the particle, while 
the two remaining sites, the lone-pair sites LP, are placed within the hard 
sphere at a distance $0.45\sigma$ from its center. Besides the hard-sphere 
interaction, preventing different particles to sample relative distances 
smaller than $\sigma$, only the H and LP sites of distinct particles interact 
via a square well (SW) potential $u_{SW}$ of width $\delta=0.15\sigma$ and 
depth $u_0$, i.e. 
\begin{equation}
u_{SW}(r)\,=\,\left\{
		\begin{array}{cc}
			-u_0 & r\,<\,\delta \\
			0 & r\,\ge\,\delta
		\end{array}
	\right.
\end{equation}
where $r$ is here the distance between H and LP sites. We remark that the 
choice $\delta=0.15 \sigma$ guarantees that multiple bonding can not take 
place at the same site.

\section{Simulations}
Three types of Monte Carlo simulations (Metropolis Grand Canonical, GCMC, 
Umbrella Sampling Grand Canonical, US-GCMC and Gibbs Ensemble GEMC) have been 
performed, with the aim of locating the critical point of the model, assessing 
the consistency with the Ising universality class and evaluating the 
gas-liquid coexistence curve. Throughout the remaining sections, we use $u_0$ 
as energy scale, $\sigma$ as length scale and reduced units in which 
$k_{\rm B}=1$, thus measuring $T$ and the chemical potential $\mu$ in units of 
$u_0/k_{\rm B}$ and $u_0$ respectively. 

We define a MC step as an average of $N_{\Delta}$ attempts to translate and 
rotate a randomly chosen particle and an average $N_{N}$ attempts to insert or 
remove a particle. The translation in each direction is uniformly chosen 
between $\pm0.05$ and the rotations are performed around a random axis of an 
angle uniformly distributed between $\pm 0.1$\,rad. Unless otherwise stated, 
$N_{\Delta}/N_N=500$. The choice of such a large ratio between 
translation/rotation and insertion/deletion attempts is dictated by the 
necessity of ensuring a proper equilibration at fixed $N$. This is 
particularly important in the case of particles with short-range and highly 
directional interactions, since the probability of inserting a particle with 
the correct orientation and position for bonding is significantly reduced as 
compared to the case of spherical interactions. 

\subsection{Critical Point Estimate and Finite Size Scaling Analysis}
\label{sec:cp}

The location of the critical point has been performed through the 
comparison of the fluctuation distribution of the ordering operator 
$\mathcal{M}$ at the critical point with the universal distribution 
characterizing the Ising class~\cite{Wilding_96}. The ordering operator 
$\mathcal{M}$ of the gas-liquid transition is a linear combination 
$\mathcal{M}\sim \rho +s u$, where $\rho$ is the number density, 
$u$ is the energy density of the system, and $s$ is the field mixing parameter. 
Exactly at the critical point, fluctuations of $\mathcal{M}$ are found to 
follow a known universal distribution, i.e. apart from a scaling factor, the 
same that characterizes the fluctuations of the magnetization in the Ising 
model~\cite{Wilding_96}. 
Finite Size Scaling (FSS) analysis has been performed to test the appartenance 
of the PMW model to the three dimensional Ising class. Recent applications of 
this method to soft matter can be found in Ref.~\cite{Miller_03,puertas,horbach}.

To locate the critical point we perform, at each size, GCMC simulations, i.e. 
at fixed $T$, $\mu$ and volume $V$. We follow the standard procedure of tuning 
$T$ and $\mu$ until the simulated system shows ample density fluctuations, 
signaling the proximity of the critical point. Once a reasonable guess of the 
critical point in the $(T,\mu)$-plane has been reached, we start at least 10 
independent GCMC simulations to improve the statistics of the fluctuations in 
the number of particles $N$ in the box and of the potential energy $E$. 

The precise evaluation of the critical temperature and chemical potential 
at all system sizes is performed by a fitting procedure associated to 
histogram reweighting. We briefly recall~\cite{histrew} that this technique 
allows us to predict, from the joint distribution $P(N,E;T,\mu)$ obtained from 
a simulation, the distribution at a different $T'$ and $\mu'$ through the 
ratio of the Boltzmann factors as follows:
\begin{equation}
\frac{P(N,E;\mu',T')}{P(N,E;\mu,T)}\, \sim \, 
 \frac{e^{-\beta'E}}{e^{-\beta E}}
  \frac{e^{\beta'\mu' N}}{e^{\beta\mu N}} \, 
   \sim \, e^{(\beta-\beta')E}e^{(\beta'\mu'-\beta\mu)N}\:. 
\end{equation}
where the proportionality constant can be calculated imposing the 
normalization of $P(N,E;\mu',T')$. The re-weighting procedure offers reliable 
results provided $T'$ and $\mu'$ are within few percents of $T$ and $\mu$.

We implement a least-square fit procedure to evaluate the values of $T$, 
$\mu$ and $s$ for which the reweighted distribution of $\mathcal{M}$ 
is closest to the known form for Ising-like systems~\cite{russi}.
The result of the fit provides the best estimate for the critical temperature 
$T_{\rm c}$, the critical chemical potential $\mu_{\rm c}$ and $s$. The 
location of the critical point has been performed for boxes of sizes $6$ 
through $9$, corresponding to an average number of particles ranging from 
about $60$ to over $400$. The simulations for larger boxes have been started 
at the critical parameters calculated for the closer smaller box. The $L=9$
simulation required about one month of CPU time for each of the 10 studied 
replicas on a 2.8\,GHz Intel Xeon.

\subsection{Phase coexistence}
\label{sec:phase}
We have calculated the phase coexistence curve using US-GCMC and GEMC simulations. 
The US-GCMC method is a natural extension of the standard GCMC 
method used to locate the critical point discussed above. Once the density 
fluctuations at the critical point have been evaluated, one can again apply the 
histogram reweigthing method to predict the shape of the density fluctuations 
at $T<T_{\rm c}$ for any fixed $\mu$ value. The value of $\mu$ for which 
coexistence between a gas and a liquid phase is present is chosen by selecting 
the value $\mu_{\rm x}$ for which the areas underneath the two peaks of 
$P({\cal M})$, the ordering operator fluctuation distribution, are equal.

We stress that performing a standard (Metropolis) GCMC simulation at a 
temperature even a few percent lower than $T_{\rm c}$ is not feasible, due to 
the large free-energy barrier separating the two phases which would prevent 
the simulated system to sample both liquid and gas configurations, thus 
yielding physically meaningless results. Several techniques have been developed 
to overcome this problem, among which the Umbrella Sampling Monte 
Carlo~\cite{umbrella} which has been used to perform the calculations 
presented herein. The US is a biased sampling technique which aims to flatten 
the free energy barrier between the two phases modifying the standard GCMC 
insertion/removal~\cite{frenkelsmith} probabilities as follows:
\begin{eqnarray}
\label{umbprob}
P_{{\rm ins}}^{{\rm US-GCMC}}\,&=\,P_{{\rm ins}}^{{\rm GCMC}}\frac{w(N)}{w(N+1)}\: \\ 
\nonumber
P_{{\rm rem}}^{{\rm US-GCMC}}\,&=\,P_{{\rm rem}}^{{\rm GCMC}}\frac{w(N)}{w(N-1)}\:.
\end{eqnarray}
In these last equations $w(N)$ is a forecast on the real $P(N)$ which 
we have repeatedly extracted from previous higher $T$ simulations through the 
histogram reweighting technique. If the predicted fluctuation distribution $w(N)$ 
is a good approximation to the real $P(N)$, the resulting biased fluctuation will 
result flat in $N$ and the system will thus not experience any difficulty in 
crossing the barrier between the liquid and gas phase. As shown in~\cite{umbrella}, 
the unbiased fluctuation distribution can be easily recovered adding a reverse 
bias to the results obtained with the insertion/removal probability in 
Eq.~\ref{umbprob}.

Starting from the critical point, we evaluate the phase diagram iterating the 
above procedure, progressively lowering $T$, down to the point where 
equilibration was not achieved any longer. Indeed, on cooling, due to the 
formation of a well-connected tetrahedral network, the dynamics in the liquid 
side slows down considerably~\cite{pwmnoi}. We have been very careful in 
progressively increasing the ratio $N_{\Delta}/N_N$ to compensate for the 
slowing down of the dynamics and the extremely slow equilibration times of the 
liquid phase. At the lower $T$, $N_{\Delta}/N_N=10000$. This sets a bound to 
the smallest $T$ which can be investigated.

We have also implemented a Gibbs Ensemble evaluation of the phase diagram. 
The GEMC method was designed~\cite{GEMC} to study 
coexistence in the region where the gas-liquid free-energy barrier is 
sufficiently high to avoid crossing between the two phases. 
Since nowadays this is a standard method in computational physics, we do 
not discuss it here and limit ourselves to noting that also in this case it is 
important to progressively increase, on cooling, the ratio $N_{\Delta}/N_N$ to 
account for the slow dynamics characterizing 
the liquid state. We have studied a system of (total) 350 particles which 
partition themselves into two boxes whose total volume is $2868\sigma^3$, 
corresponding to an average density of $\rho=0.122$. At the lowest $T$ this 
corresponds to roughly $320$ particles in the liquid box (of side $\approx 8\sigma$) 
and about 30 particles in the gas box (of side $\approx 13\sigma$). Equilibration 
at the lowest reported $T$ required about three months of computer time.

\begin{figure}[tb]
\includegraphics[width=8cm,clip=true]{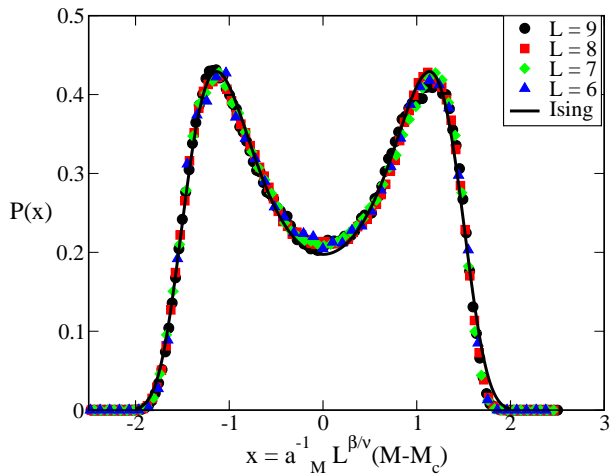}
\caption{Distributions of the fluctuations of the ordering operator for all studied sizes at the apparent critical point. The 
factor $a_{\mathcal{M}}^{-1}=0.34$ has been chosen to scale $P(x)$ to unit variance. 
The theoretical curve for the Ising model (full line) is from~\cite{russi}.}
\label{fig:pdix}
\end{figure}

\section{Results}
\begin{figure}[tb]
\includegraphics[width=8cm,clip=true]{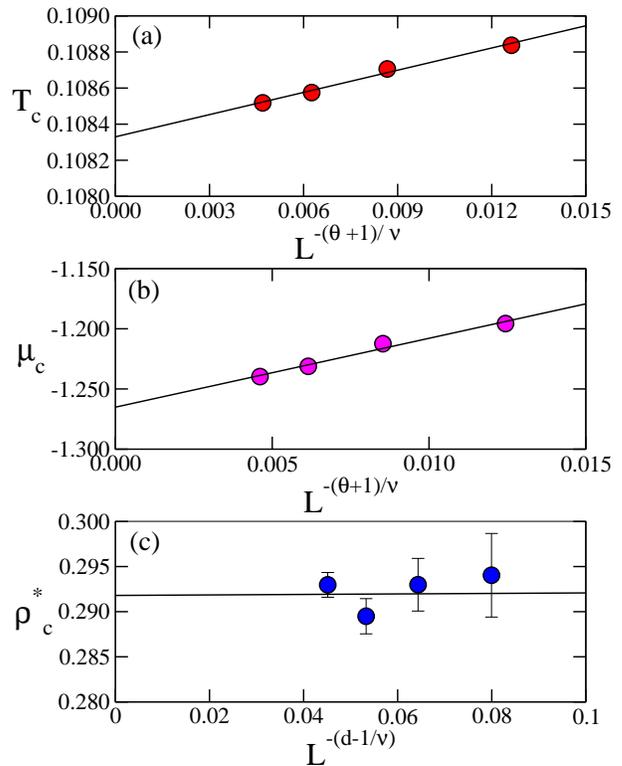}
\caption{Size dependence of the apparent critical temperature 
$T_{\rm c}(L)$ (a), the apparent critical chemical potential $\mu_{\rm c}(L)$ (b) 
and the density at the true critical point $\rho_{\rm c}^{*}(L)$ (c).}
\label{fig:pc}
\end{figure}

We start showing the distributions of the ordering operator fluctuations at 
the (apparent) critical points for all studied sizes. Implementing the fitting 
procedure described in Sec.~\ref{sec:cp}, using histogram reweighting, we 
first evaluate the values of the critical parameters and the shape of the 
density fluctuations at the critical point. The resulting distributions, for 
all investigated $L$, are shown as a function of the scaled variable 
$x \equiv a_M^{-1} L^{\beta/\nu} ({\cal M-M}_{\rm c})$ in Fig.~\ref{fig:pdix}. 
Here $\nu$ is the critical exponent of the correlation length and $\beta$ is 
critical exponent of the order parameter. Within the $d=3$ Ising universality 
class $\nu=0.629$ and $\beta=0.326$. From the fits we find that the 
non-universal amplitude is $a_M^{-1}=0.34$ (independent on $L$). The size 
dependence of $T_{\rm c}$ and $\mu_{\rm c}$ for $L=6$ to $L=9$ is shown in 
Fig.~\ref{fig:pc}. Finite size scaling predicts $T_{\rm c} \sim 
L^{-(\theta+1)/\nu}$ and $\mu_{\rm c} \sim L^{-(\theta+1)/\nu}$, where 
$\theta=0.54$ (Ising) is the universal correction to the scaling 
exponent~\cite{Wilding_96}. Fig.~\ref{fig:pc}(a-b) shows that the size 
dependence of the critical parameters is consistent with the 
expected universality class. Extrapolating the observed size dependence to 
$L \rightarrow \infty$ it is possible to provide an estimate for the bulk 
behavior of the PMW potential. We find $T_{\rm c}^{bulk}=0.1083(3)$, 
$\mu_{\rm c}^{bulk}=-1.265(1)$.
Fig.~\ref{fig:pc}(c) shows the $L$ dependence of $\rho_{\rm c}^{*}$, the 
latter being the system density (evaluated via histogram reweighting for 
each size) at $T_{\rm c}^{bulk}$ and $\mu_{\rm c}^{bulk}$. Finite size 
scaling predicts $\rho_{\rm c}^{*}\sim L^{-(d-1/\nu)}$. The resulting $L$ 
dependence of $\rho_{\rm c}^*$ is consistent with the theoretical 
prediction, despite the fact that $\rho_{\rm c}^*$ data are the noisiest, 
since the estimate of density is the most delicate and the error is at 
least of the order of one particle over the volume. The field mixing parameter 
$s$ has been difficult to infer precisely due to its small value and to the 
discrete nature of the square well interactions which makes the distribution 
of $N$ and $E$ quantized over integer numbers. A value of $s \approx 0.07$, 
consistent with the results reported in~\cite{bianchiprl} for a similar model, 
allows a simultaneous fit of all distribution functions, independently from $L$. 

\begin{figure}[tb]
\includegraphics[width=8cm,clip=true]{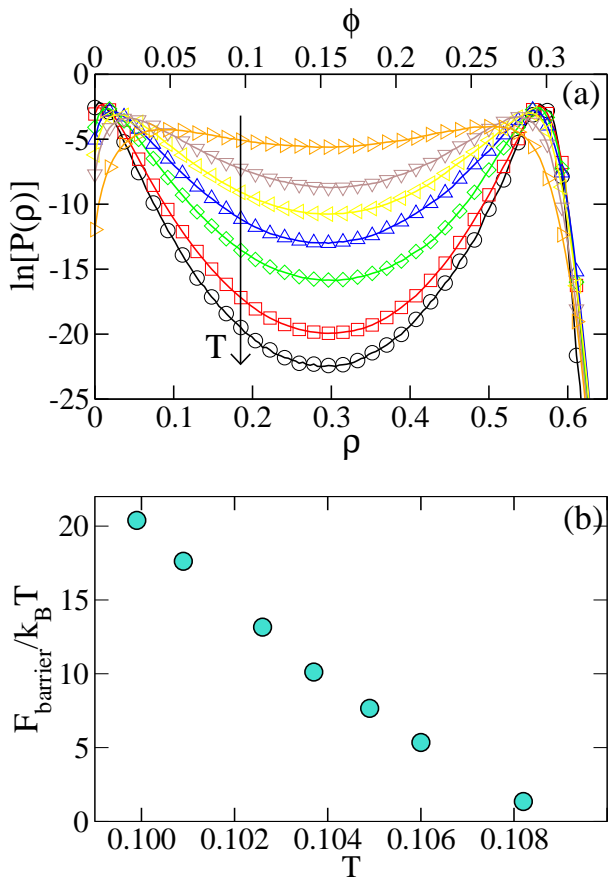}
\caption{{\it (a)}: Logarithm of the distribution of density fluctuations 
$P(\rho)$ at $T~=~0.1082$, $0.1060$, $0.1049$, $0.1037$, $0.1026$, 
$0.1009$, $0.0999$, providing the negative of the density dependence of the 
free energy. The difference between the value of $P(\rho)$ at the valley and 
at the top is a measure of the free energy barrier separating the gas and the 
liquid state {\it (b)}: Free energy barrier (in units of $k_{\rm B}T$) between 
gas and liquid versus $T$ at coexistence.}
\label{fig:logPrho}
\end{figure}

\begin{figure}[tb]
\begin{center}
\includegraphics[width=8cm,clip=true]{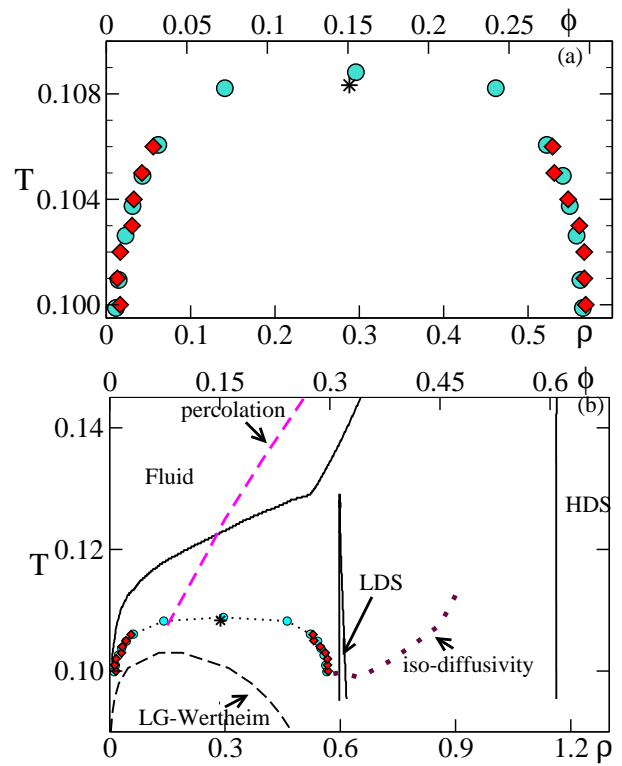}
\end{center}
\caption{{\it (a)}: Calculated gas-liquid coexistence for the PMW. Both GEMC (diamonds) 
and US-GCMC (circles) simulations results are shown. The star indicates the 
bulk ($L \rightarrow \infty$) critical point estimate. {\it (b)}: Extended phase diagram 
for the PMW, including: (i) the theoretical (Wertheim) gas-liquid coexistence line (dashed), 
(ii) the field of stability of the fluid, of the high-density crystal (HDS) and of the 
low-density crystal (LDS) phases from Ref.~\cite{Veg98a}; (iii) an iso-diffusivity line 
(for the smallest investigated value of $D$) from Ref.~\cite{pwmnoi}; (iv) percolation 
line from Ref.~\cite{pwmnoi}.}
\label{fig:coex}
\end{figure}

Next we discuss the gas-liquid coexistence curve for the PMW. Since only 
close to the critical point size effects are relevant, we have performed the 
calculations at $L=6$. Fig.~\ref{fig:logPrho} shows the behavior of the 
density fluctuations $P(\rho)$ (in log scale), evaluated with US-GCMC 
simulations (see Sec.~\ref{sec:phase}) along the coexistence curve. The 
difference between the peak and the valley in $\ln[P(\rho)]$ is a measure, 
in unit of $k_{\rm B}T$, of the activation free-energy 
$F_{barrier}/k_{\rm B}T$ needed to cross from the gas to the liquid phase 
and vice-versa. At the lowest studied temperature, the barrier reaches a 
value of about $20k_{\rm B}T$, which would clearly be impossible to 
overcome without the use of a biased sampling technique. Fig.~\ref{fig:logPrho} 
also shows the $T$ dependence of the barrier height, which indeed becomes of 
the order of the thermal energy close to the critical point.

The phase diagram resulting from our calculations is reported in 
Fig.~\ref{fig:coex}-(a). Both US-GCMC and GEMC data are reported, showing a 
perfect agreement for all studied $T$ proving that, despite the long 
equilibration times required, an accurate determination of the phase diagram 
for this model of patchy particles can nowadays be achieved. 
Fig.~\ref{fig:coex}-(b) shows the same data, together with the fluid-crystal 
coexistence lines calculated by Vega and Monson~\cite{Veg98a} and complemented 
with the bond percolation line from Ref.~\cite{pwmnoi}. We also report in the graph 
a so-called iso-diffusivity line, i.e. the set of points in the $T-\rho$ plane
in which the diffusion coefficient $D$ is constant. By selecting the smallest value
of $D$ which can be calculated with the present time computational facilities,
the iso-diffusivity line provides an estimate of the shape of the glass line.
Fig.~\ref{fig:coex} also shows the coexistence curve evaluated from the Wertheim theory, from Ref.~\cite{Veg98a}. 

Several observations arise from the data in Fig.~\ref{fig:coex}. 
First of all, we observe that the critical point, as well as the 
liquid branch of the coexistence curve, are characterized by 
a percolating (but transient) structure of bonds between pairs of H and LP. 
This is not unexpected, since the propagation of infinite range 
correlations, characteristic of a critical point, does require the 
presence of a spanning cluster~\cite{coniglio}. 
Particle diffusion is still significant, despite the presence of the transient 
percolating network of bonds, as shown by the comparison between the
phase-coexistence and the small $D$ iso-diffusivity line. In the region of $T$ 
where dynamics becomes so slow that equilibration can not be achieved on the 
present computational time scale, it becomes impossible also to evaluate the 
gas-liquid coexistence.

The gas-liquid coexistence is found to be metastable respect to fluid-crystal 
coexistence, in analogy with the case of particles interacting through 
spherical short-range potentials. This notwithstanding and differently from 
the case of spherical interacting particles, we never observe crystallization 
close to the critical point. This suggests that the increase in local density 
brought in by the critical fluctuations does not sufficiently couple with the 
orientational ordering required for the formation of the open diamond crystal 
structure.

The comparison between the GEMC and US-GCMC simulation phase diagram with the 
theoretical predictions of the Wertheim theory suggests
that the latter provides a quite accurate estimate for $T_{\rm c}$, whereas the 
$\rho$ dependence is only approximate. Indeed, the Wertheim theory predicts 
a vapor-liquid critical point at $T_{\rm c}=0.1031$ and 
$\rho_{\rm c}=0.279$~\cite{Veg98a}. 
These differences between theoretical predictions and numerical data 
confirm the conclusions that have been previously reached for models 
of patchy particles with different number of sticky spots~\cite{bianchiprl}. 

Finally, we note that in a very limited $T$-interval (less than 10 per 
cent of $T_{\rm c}$), the liquid density approximatively reaches a 
value comparable to the diamond crystal density, which for the present model has 
been calculated~\cite{Veg98a} $0.4880<\rho<0.6495$. Thus, as for the crystal state, 
the density of the liquid phase at coexistence is rather small, 
approximatively a factor of two smaller as compared to the case of spherically 
interacting particles. 

\begin{figure}[tb]
\includegraphics[width=8cm,clip=true]{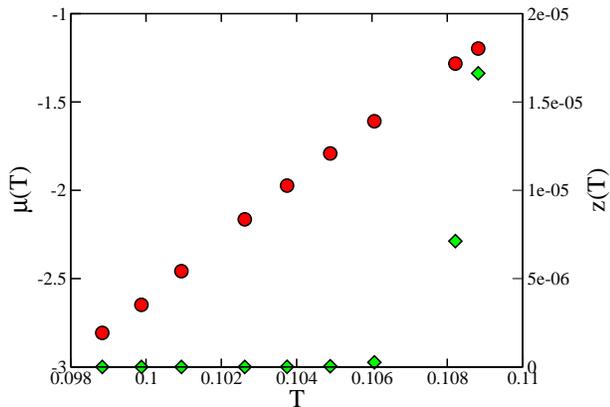}
\caption{Chemical potential (circles) and fugacity $z \equiv e^{\beta\mu}$ 
(diamonds) versus temperature $T$ at coexistence for $L=6$.}
\label{fig:muT}
\end{figure}

Finally, for the sake of completeness we show in Fig.~\ref{fig:muT} the 
location of the gas-liquid coexistence in the $\mu-T$ plane. 

\section{Conclusions}
In this article we have presented an accurate determination of the 
critical point and of the gas-liquid phase coexistence curve for a 
primitive model for water, introduced by Kolafa and Nezbeda~\cite{Kol87a}. 
Despite its original motivation, the PMW can also be studied as 
a model for patchy colloidal particles and, perhaps, as an elementary 
model for describing patchy interactions in proteins. To this extent, 
it is particularly important to understand the qualitative features of 
the phase diagram, the stability or metastability of the gas-liquid line and 
the propensity to crystallize.

We have shown that the critical fluctuations are consistent with the Ising 
universality class, both via the analysis of the shape of the density 
fluctuation distribution and via the size dependence of the critical 
parameters. The determination of the critical point allows us to prove that, for 
the present model, the liquid phase is not present in equilibrium, and it is 
only observed in a metastable condition. In the case of spherical interaction 
potentials, the thermodynamic stability of the gas-liquid critical point 
is achieved when the range of the interaction potential becomes of the order 
of one tenth of the particle diameter~\cite{Vli00a}. Data reported in this 
article (see Fig.~\ref{fig:coex}) confirm that the property of short-range 
potentials of missing a proper equilibrium liquid phase is retained in 
short-range patchy particles. It would be of interest to study for which 
critical value of the interaction range a proper liquid phase appears in 
such patchy particles.

The evaluation of the gas-liquid coexistence has been performed using two 
distinct methods, the US-GCMC and the GEMC, and a very good agreement has been 
recorded despite the difficulties in equilibrating the liquid phase at low $T$.
The present results, together with previous studies of the crystal phase and of 
the slow low-$T$ dynamics, offer a coherent picture of the behavior of the 
model. It is shown that phase separation is intervening only at low $T$ for 
$\rho \lesssim 0.6$, a value significantly smaller than the corresponding 
value for spherical interaction potentials. This is consistent with the fact 
that the number of attractive interactions in which a particle can be engaged 
(four in the present model) controls the minimum density of the liquid phase. 
These results elucidate the different nature of bonded liquids (the so-called 
network forming) with respect to the category of spherically interacting 
liquids. For bonded systems, there is an intermediate region, between the high 
density packed structure and the unstable region, in which gas-liquid phase 
separation is observed, which is not accessible to spherically interacting 
particles. In this intermediate density region, the system is structured in a 
percolating network and both static and dynamic quantities are controlled by 
the presence of bonds. This picture, which has been developing in a progression 
of studies~\cite{Zac05a,pwmnoi,silica,valency}, is confirmed by the present 
results.

\section*{Acknowledgments}

We acknowledge support from MIUR-Prin and
MCRTN-CT-2003-504712.

\section*{References}
\bibliographystyle{apsrev}
\bibliography{pmw}

\begin{thebibliography}{34}
\expandafter\ifx\csname natexlab\endcsname\relax\def\natexlab#1{#1}\fi
\expandafter\ifx\csname bibnamefont\endcsname\relax
  \def\bibnamefont#1{#1}\fi
\expandafter\ifx\csname bibfnamefont\endcsname\relax
  \def\bibfnamefont#1{#1}\fi
\expandafter\ifx\csname citenamefont\endcsname\relax
  \def\citenamefont#1{#1}\fi
\expandafter\ifx\csname url\endcsname\relax
  \def\url#1{\texttt{#1}}\fi
\expandafter\ifx\csname urlprefix\endcsname\relax\def\urlprefix{URL }\fi
\providecommand{\bibinfo}[2]{#2}
\providecommand{\eprint}[2][]{\url{#2}}

\bibitem[{\citenamefont{Kolafa and Nezbeda}(1987)}]{Kol87a}
\bibinfo{author}{\bibfnamefont{J.}~\bibnamefont{Kolafa}} \bibnamefont{and}
  \bibinfo{author}{\bibfnamefont{I.}~\bibnamefont{Nezbeda}},
  \bibinfo{journal}{Mol. Phys.} \textbf{\bibinfo{volume}{61}},
  \bibinfo{pages}{161} (\bibinfo{year}{1987}).

\bibitem[{\citenamefont{Wertheim}(1984{\natexlab{a}})}]{Wer84a}
\bibinfo{author}{\bibfnamefont{M.~S.} \bibnamefont{Wertheim}},
  \bibinfo{journal}{J. Stat. Phys.} \textbf{\bibinfo{volume}{35}},
  \bibinfo{pages}{19} (\bibinfo{year}{1984}{\natexlab{a}}).

\bibitem[{\citenamefont{Wertheim}(1984{\natexlab{b}})}]{Wer84b}
\bibinfo{author}{\bibfnamefont{M.~S.} \bibnamefont{Wertheim}},
  \bibinfo{journal}{J. Stat. Phys.} \textbf{\bibinfo{volume}{35}},
  \bibinfo{pages}{35} (\bibinfo{year}{1984}{\natexlab{b}}).

\bibitem[{\citenamefont{Ghonasci and Chapman}(1993)}]{Gho93a}
\bibinfo{author}{\bibfnamefont{D.}~\bibnamefont{Ghonasci}} \bibnamefont{and}
  \bibinfo{author}{\bibfnamefont{W.~G.} \bibnamefont{Chapman}},
  \bibinfo{journal}{Mol. Phys.} \textbf{\bibinfo{volume}{79}},
  \bibinfo{pages}{291} (\bibinfo{year}{1993}).

\bibitem[{\citenamefont{{Sear} and {Jackson}}(1996)}]{Sea96a}
\bibinfo{author}{\bibfnamefont{R.~P.} \bibnamefont{{Sear}}} \bibnamefont{and}
  \bibinfo{author}{\bibfnamefont{G.}~\bibnamefont{{Jackson}}},
  \bibinfo{journal}{J. Chem. Phys.} \textbf{\bibinfo{volume}{105}},
  \bibinfo{pages}{1113} (\bibinfo{year}{1996}).

\bibitem[{\citenamefont{{Duda} et~al.}(1998)\citenamefont{{Duda}, {Segura},
  {Vakarin}, {Holovko}, and {Chapman}}}]{Dud98a}
\bibinfo{author}{\bibfnamefont{Y.}~\bibnamefont{{Duda}}},
  \bibinfo{author}{\bibfnamefont{C.~J.} \bibnamefont{{Segura}}},
  \bibinfo{author}{\bibfnamefont{E.}~\bibnamefont{{Vakarin}}},
  \bibinfo{author}{\bibfnamefont{M.~F.} \bibnamefont{{Holovko}}},
  \bibnamefont{and} \bibinfo{author}{\bibfnamefont{W.~G.}
  \bibnamefont{{Chapman}}}, \bibinfo{journal}{J. Chem. Phys.}
  \textbf{\bibinfo{volume}{108}}, \bibinfo{pages}{9168} (\bibinfo{year}{1998}).

\bibitem[{\citenamefont{Peery and Evans}(2003)}]{Pee03a}
\bibinfo{author}{\bibfnamefont{T.~B.} \bibnamefont{Peery}} \bibnamefont{and}
  \bibinfo{author}{\bibfnamefont{G.~T.} \bibnamefont{Evans}},
  \bibinfo{journal}{J. Chem Phys.} \textbf{\bibinfo{volume}{118}},
  \bibinfo{pages}{2286} (\bibinfo{year}{2003}).

\bibitem[{\citenamefont{Kalyuzhnyi and Cummings}(2003)}]{Kal03a}
\bibinfo{author}{\bibfnamefont{Y.~V.} \bibnamefont{Kalyuzhnyi}}
  \bibnamefont{and} \bibinfo{author}{\bibfnamefont{P.~T.}
  \bibnamefont{Cummings}}, \bibinfo{journal}{J. Chem Phys.}
  \textbf{\bibinfo{volume}{118}}, \bibinfo{pages}{6437} (\bibinfo{year}{2003}).

\bibitem[{\citenamefont{Nezbeda et~al.}(1989)\citenamefont{Nezbeda, Kolafa, and
  Kalyuzhnyi}}]{Nez89a}
\bibinfo{author}{\bibfnamefont{I.}~\bibnamefont{Nezbeda}},
  \bibinfo{author}{\bibfnamefont{J.}~\bibnamefont{Kolafa}}, \bibnamefont{and}
  \bibinfo{author}{\bibfnamefont{Y.}~\bibnamefont{Kalyuzhnyi}},
  \bibinfo{journal}{Mol. Phys.} \textbf{\bibinfo{volume}{68}},
  \bibinfo{pages}{143} (\bibinfo{year}{1989}).

\bibitem[{\citenamefont{Nezbeda and Iglesias-Silva}(1990)}]{Nez90a}
\bibinfo{author}{\bibfnamefont{I.}~\bibnamefont{Nezbeda}} \bibnamefont{and}
  \bibinfo{author}{\bibfnamefont{G.}~\bibnamefont{Iglesias-Silva}},
  \bibinfo{journal}{Mol. Phys.} \textbf{\bibinfo{volume}{69}},
  \bibinfo{pages}{767} (\bibinfo{year}{1990}).

\bibitem[{\citenamefont{Vega and Monson}(1998)}]{Veg98a}
\bibinfo{author}{\bibfnamefont{C.}~\bibnamefont{Vega}} \bibnamefont{and}
  \bibinfo{author}{\bibfnamefont{P.~A.} \bibnamefont{Monson}},
  \bibinfo{journal}{J. Chem. Phys.} \textbf{\bibinfo{volume}{109}},
  \bibinfo{pages}{9938} (\bibinfo{year}{1998}).

\bibitem[{\citenamefont{DeMichele et~al.}(2006)\citenamefont{DeMichele,
  Gabrielli, Tartaglia, and Sciortino}}]{pwmnoi}
\bibinfo{author}{\bibfnamefont{C.}~\bibnamefont{DeMichele}},
  \bibinfo{author}{\bibfnamefont{S.}~\bibnamefont{Gabrielli}},
  \bibinfo{author}{\bibfnamefont{P.}~\bibnamefont{Tartaglia}},
  \bibnamefont{and}
  \bibinfo{author}{\bibfnamefont{F.}~\bibnamefont{Sciortino}},
  \bibinfo{journal}{Journal of Physical Chemistry B}
  \textbf{\bibinfo{volume}{110}}, \bibinfo{pages}{8064} (\bibinfo{year}{2006}),
  ISSN \bibinfo{issn}{1520-6106},
  \urlprefix\url{http://pubs3.acs.org/acs/journals/doilookup?in_doi=10.1021/jp%
056380y}.

\bibitem[{\citenamefont{Lomakin et~al.}(1999)\citenamefont{Lomakin, Asherie,
  and Benedek}}]{Lom99a}
\bibinfo{author}{\bibfnamefont{A.}~\bibnamefont{Lomakin}},
  \bibinfo{author}{\bibfnamefont{N.}~\bibnamefont{Asherie}}, \bibnamefont{and}
  \bibinfo{author}{\bibfnamefont{G.~B.} \bibnamefont{Benedek}},
  \bibinfo{journal}{Proc. Natl. Acad. Sci.} \textbf{\bibinfo{volume}{96}},
  \bibinfo{pages}{9465} (\bibinfo{year}{1999}).

\bibitem[{\citenamefont{Sear}(1999)}]{Sea99c}
\bibinfo{author}{\bibfnamefont{R.~P.} \bibnamefont{Sear}}, \bibinfo{journal}{J.
  Chem. Phys.} \textbf{\bibinfo{volume}{111}}, \bibinfo{pages}{4800}
  (\bibinfo{year}{1999}).

\bibitem[{\citenamefont{Kern and Frenkel}(2003)}]{Ker03a}
\bibinfo{author}{\bibfnamefont{N.}~\bibnamefont{Kern}} \bibnamefont{and}
  \bibinfo{author}{\bibfnamefont{D.}~\bibnamefont{Frenkel}},
  \bibinfo{journal}{J.Chem Phys.} \textbf{\bibinfo{volume}{118}},
  \bibinfo{pages}{9882} (\bibinfo{year}{2003}).

\bibitem[{\citenamefont{Doye et~al.}(2007)\citenamefont{Doye, Louis, Lin,
  Allen, Noya, Wilber, Kok, and Lyus}}]{doye}
\bibinfo{author}{\bibfnamefont{J.~P.~K.} \bibnamefont{Doye}},
  \bibinfo{author}{\bibfnamefont{A.~A.} \bibnamefont{Louis}},
  \bibinfo{author}{\bibfnamefont{I.-C.} \bibnamefont{Lin}},
  \bibinfo{author}{\bibfnamefont{L.~R.} \bibnamefont{Allen}},
  \bibinfo{author}{\bibfnamefont{E.~G.} \bibnamefont{Noya}},
  \bibinfo{author}{\bibfnamefont{A.~W.} \bibnamefont{Wilber}},
  \bibinfo{author}{\bibfnamefont{H.~C.} \bibnamefont{Kok}}, \bibnamefont{and}
  \bibinfo{author}{\bibfnamefont{R.}~\bibnamefont{Lyus}},
  \emph{\bibinfo{title}{Controlling crystallization and its absence: Proteins,
  colloids and patchy models}} (\bibinfo{year}{2007}),
  \eprint{cond-mat/0701074},
  \urlprefix\url{http://www.citebase.org/abstract?id=oai:arXiv.org:cond-mat/07%
01074}.

\bibitem[{\citenamefont{Manoharan et~al.}(2003)\citenamefont{Manoharan,
  Elsesser, and Pine}}]{Man03a}
\bibinfo{author}{\bibfnamefont{V.~N.} \bibnamefont{Manoharan}},
  \bibinfo{author}{\bibfnamefont{M.~T.} \bibnamefont{Elsesser}},
  \bibnamefont{and} \bibinfo{author}{\bibfnamefont{D.~J.} \bibnamefont{Pine}},
  \bibinfo{journal}{Science} \textbf{\bibinfo{volume}{301}},
  \bibinfo{pages}{483} (\bibinfo{year}{2003}).

\bibitem[{\citenamefont{Bianchi et~al.}(2006)\citenamefont{Bianchi, Largo,
  Tartaglia, Zaccarelli, and Sciortino}}]{bianchiprl}
\bibinfo{author}{\bibfnamefont{E.}~\bibnamefont{Bianchi}},
  \bibinfo{author}{\bibfnamefont{J.}~\bibnamefont{Largo}},
  \bibinfo{author}{\bibfnamefont{P.}~\bibnamefont{Tartaglia}},
  \bibinfo{author}{\bibfnamefont{E.}~\bibnamefont{Zaccarelli}},
  \bibnamefont{and}
  \bibinfo{author}{\bibfnamefont{F.}~\bibnamefont{Sciortino}},
  \bibinfo{journal}{Physical Review Letters} \textbf{\bibinfo{volume}{97}},
  \bibinfo{eid}{168301} (pages~\bibinfo{numpages}{4}) (\bibinfo{year}{2006}),
  \urlprefix\url{http://link.aps.org/abstract/PRL/v97/e168301}.

\bibitem[{\citenamefont{Zaccarelli et~al.}(2005)\citenamefont{Zaccarelli,
  Buldyrev, Nave, Moreno, Saika-Voivod, Sciortino, and Tartaglia}}]{Zac05a}
\bibinfo{author}{\bibfnamefont{E.}~\bibnamefont{Zaccarelli}},
  \bibinfo{author}{\bibfnamefont{S.~V.} \bibnamefont{Buldyrev}},
  \bibinfo{author}{\bibfnamefont{E.~L.} \bibnamefont{Nave}},
  \bibinfo{author}{\bibfnamefont{A.~J.} \bibnamefont{Moreno}},
  \bibinfo{author}{\bibfnamefont{I.}~\bibnamefont{Saika-Voivod}},
  \bibinfo{author}{\bibfnamefont{F.}~\bibnamefont{Sciortino}},
  \bibnamefont{and}
  \bibinfo{author}{\bibfnamefont{P.}~\bibnamefont{Tartaglia}},
  \bibinfo{journal}{Phys. Rev Lett.} \textbf{\bibinfo{volume}{94}},
  \bibinfo{pages}{218301} (\bibinfo{year}{2005}).

\bibitem[{\citenamefont{Sciortino et~al.}(2005)\citenamefont{Sciortino,
  Buldyrev, {De Michele}, Ghofraniha, {La Nave}, Moreno, Mossa, Tartaglia, and
  Zaccarelli}}]{Sci04bCPC}
\bibinfo{author}{\bibfnamefont{F.}~\bibnamefont{Sciortino}},
  \bibinfo{author}{\bibfnamefont{S.}~\bibnamefont{Buldyrev}},
  \bibinfo{author}{\bibfnamefont{C.}~\bibnamefont{{De Michele}}},
  \bibinfo{author}{\bibfnamefont{N.}~\bibnamefont{Ghofraniha}},
  \bibinfo{author}{\bibfnamefont{E.}~\bibnamefont{{La Nave}}},
  \bibinfo{author}{\bibfnamefont{A.}~\bibnamefont{Moreno}},
  \bibinfo{author}{\bibfnamefont{S.}~\bibnamefont{Mossa}},
  \bibinfo{author}{\bibfnamefont{P.}~\bibnamefont{Tartaglia}},
  \bibnamefont{and}
  \bibinfo{author}{\bibfnamefont{E.}~\bibnamefont{Zaccarelli}},
  \bibinfo{journal}{Comp. Phys. Comm.} \textbf{\bibinfo{volume}{169}},
  \bibinfo{pages}{166} (\bibinfo{year}{2005}).

\bibitem[{\citenamefont{{Zaccarelli} et~al.}(2006)\citenamefont{{Zaccarelli},
  {Saika-Voivod}, {Buldyrev}, {Moreno}, {Tartaglia}, and
  {Sciortino}}}]{zaccalungo}
\bibinfo{author}{\bibfnamefont{E.}~\bibnamefont{{Zaccarelli}}},
  \bibinfo{author}{\bibfnamefont{I.}~\bibnamefont{{Saika-Voivod}}},
  \bibinfo{author}{\bibfnamefont{S.~V.} \bibnamefont{{Buldyrev}}},
  \bibinfo{author}{\bibfnamefont{A.~J.} \bibnamefont{{Moreno}}},
  \bibinfo{author}{\bibfnamefont{P.}~\bibnamefont{{Tartaglia}}},
  \bibnamefont{and}
  \bibinfo{author}{\bibfnamefont{F.}~\bibnamefont{{Sciortino}}},
  \bibinfo{journal}{J. Chem. Phys.} \textbf{\bibinfo{volume}{124}},
  \bibinfo{pages}{4908} (\bibinfo{year}{2006}), \eprint{cond-mat/0511433}.

\bibitem[{\citenamefont{{de Michele} et~al.}(2006)\citenamefont{{de Michele},
  {Tartaglia}, and {Sciortino}}}]{silica}
\bibinfo{author}{\bibfnamefont{C.}~\bibnamefont{{de Michele}}},
  \bibinfo{author}{\bibfnamefont{P.}~\bibnamefont{{Tartaglia}}},
  \bibnamefont{and}
  \bibinfo{author}{\bibfnamefont{F.}~\bibnamefont{{Sciortino}}},
  \bibinfo{journal}{J. Chem. Phys.} \textbf{\bibinfo{volume}{125}},
  \bibinfo{pages}{4710} (\bibinfo{year}{2006}), \eprint{cond-mat/0606618}.

\bibitem[{\citenamefont{Wilding}(1996)}]{Wilding_96}
\bibinfo{author}{\bibfnamefont{N.~B.} \bibnamefont{Wilding}},
  \bibinfo{journal}{J. Phys. Cond. Matt.} \textbf{\bibinfo{volume}{9}},
  \bibinfo{pages}{585} (\bibinfo{year}{1996}).

\bibitem[{\citenamefont{Miller and Frenkel}(2003)}]{Miller_03}
\bibinfo{author}{\bibfnamefont{M.~A.} \bibnamefont{Miller}} \bibnamefont{and}
  \bibinfo{author}{\bibfnamefont{D.}~\bibnamefont{Frenkel}},
  \bibinfo{journal}{Phys. Rev. Lett.} \textbf{\bibinfo{volume}{{\bf 90}}},
  \bibinfo{pages}{135702} (\bibinfo{year}{2003}).

\bibitem[{\citenamefont{{Caballero} and {\it et al.}}(2004)}]{puertas}
\bibinfo{author}{\bibfnamefont{J.~B.} \bibnamefont{{Caballero}}}
  \bibnamefont{and} \bibinfo{author}{\bibnamefont{{\it et al.}}},
  \bibinfo{journal}{J. Chem. Phys.} \textbf{\bibinfo{volume}{121}},
  \bibinfo{pages}{2428} (\bibinfo{year}{2004}).

\bibitem[{\citenamefont{{Vink} and {Horbach}}(2004)}]{horbach}
\bibinfo{author}{\bibfnamefont{R.~L.~C.} \bibnamefont{{Vink}}}
  \bibnamefont{and}
  \bibinfo{author}{\bibfnamefont{J.}~\bibnamefont{{Horbach}}},
  \bibinfo{journal}{J. Chem. Phys.} \textbf{\bibinfo{volume}{121}},
  \bibinfo{pages}{3253} (\bibinfo{year}{2004}).

\bibitem[{\citenamefont{Ferrenberg and Swendsen}(1988)}]{histrew}
\bibinfo{author}{\bibfnamefont{A.~M.} \bibnamefont{Ferrenberg}}
  \bibnamefont{and} \bibinfo{author}{\bibfnamefont{R.~H.}
  \bibnamefont{Swendsen}}, \bibinfo{journal}{Phys. Rev. Lett.}
  \textbf{\bibinfo{volume}{61}}, \bibinfo{pages}{2635} (\bibinfo{year}{1988}).

\bibitem[{\citenamefont{Tsypin and Bl\"ote}(2000)}]{russi}
\bibinfo{author}{\bibfnamefont{M.~M.} \bibnamefont{Tsypin}} \bibnamefont{and}
  \bibinfo{author}{\bibfnamefont{H.~W.~J.} \bibnamefont{Bl\"ote}},
  \bibinfo{journal}{Phys. Rev. E} \textbf{\bibinfo{volume}{62}},
  \bibinfo{pages}{73} (\bibinfo{year}{2000}).

\bibitem[{\citenamefont{Patey and Valleau}(1975)}]{umbrella}
\bibinfo{author}{\bibfnamefont{G.~N.} \bibnamefont{Patey}} \bibnamefont{and}
  \bibinfo{author}{\bibfnamefont{J.~P.} \bibnamefont{Valleau}},
  \bibinfo{journal}{J. Chem. Phys.} \textbf{\bibinfo{volume}{63}},
  \bibinfo{pages}{2334} (\bibinfo{year}{1975}).

\bibitem[{\citenamefont{Smith and Frenkel}(1996)}]{frenkelsmith}
\bibinfo{author}{\bibfnamefont{B.}~\bibnamefont{Smith}} \bibnamefont{and}
  \bibinfo{author}{\bibfnamefont{D.}~\bibnamefont{Frenkel}},
  \emph{\bibinfo{title}{Understanding molecular simulations}}
  (\bibinfo{publisher}{Academic}, \bibinfo{address}{New York},
  \bibinfo{year}{1996}).

\bibitem[{\citenamefont{Panagiotopoulos}(1987)}]{GEMC}
\bibinfo{author}{\bibfnamefont{A.~Z.} \bibnamefont{Panagiotopoulos}},
  \bibinfo{journal}{Mol. Phys.} \textbf{\bibinfo{volume}{61}},
  \bibinfo{pages}{813} (\bibinfo{year}{1987}).

\bibitem[{\citenamefont{Coniglio and Klein}(1980)}]{coniglio}
\bibinfo{author}{\bibfnamefont{A.}~\bibnamefont{Coniglio}} \bibnamefont{and}
  \bibinfo{author}{\bibfnamefont{W.}~\bibnamefont{Klein}}, \bibinfo{journal}{J.
  Phys. A} \textbf{\bibinfo{volume}{13}}, \bibinfo{pages}{2775}
  (\bibinfo{year}{1980}).

\bibitem[{\citenamefont{Vliegenthart and Lekkerkerker}(2000)}]{Vli00a}
\bibinfo{author}{\bibfnamefont{G.~A.} \bibnamefont{Vliegenthart}}
  \bibnamefont{and} \bibinfo{author}{\bibfnamefont{H.~N.~W.}
  \bibnamefont{Lekkerkerker}}, \bibinfo{journal}{J. Chem. Phys.}
  \textbf{\bibinfo{volume}{12}}, \bibinfo{pages}{5364} (\bibinfo{year}{2000}).

\bibitem[{\citenamefont{{Sastry} et~al.}(2006)\citenamefont{{Sastry}, {La
  Nave}, and {Sciortino}}}]{valency}
\bibinfo{author}{\bibfnamefont{S.}~\bibnamefont{{Sastry}}},
  \bibinfo{author}{\bibfnamefont{E.}~\bibnamefont{{La Nave}}},
  \bibnamefont{and}
  \bibinfo{author}{\bibfnamefont{F.}~\bibnamefont{{Sciortino}}},
  \bibinfo{journal}{Journal of Statistical Mechanics: Theory and Experiment}
  \textbf{\bibinfo{volume}{12}}, \bibinfo{pages}{10} (\bibinfo{year}{2006}),
  \eprint{cond-mat/0609388}.

\end{thebibliography}

\end{document}